\documentclass[]{article}

\usepackage{amssymb,latexsym,amsmath}     
\usepackage[utf8]{inputenc} 
\usepackage[T1]{fontenc}    
\usepackage{hyperref}       
\usepackage{url}            
\usepackage{booktabs}       
\usepackage{amsfonts}       
\usepackage{nicefrac}       
\usepackage{microtype}      
\usepackage{lipsum}
\usepackage{fancyhdr}       
\usepackage{graphicx}       
\usepackage{array}          
\usepackage{tabularx}       
\graphicspath{{media/}}     

\title{On the Photoluminescence Theory in Nanocrystalline Silicon: A New Improvement

}

\author{
  Arturo Ramirez-Porras \\
  (Centro de Investigación en Ciencia \\
  e Ingeniería de Materiales - CICIMA - \\
  and Escuela de Física \\
  Universidad de Costa Rica \\
  San José, Costa Rica)\\
    \texttt{arturo.ramirez@ucr.ac.cr} \\
}

\begin{document}
\maketitle

\begin{abstract}
Silicon has long been the foundational semiconductor material for a broad range of electronic devices, owing to its numerous advantages: wide natural availability, ease of synthesis in both crystalline and amorphous forms, and relatively low production cost. However, despite these benefits, silicon is inherently limited in the field of optoelectronics due to its indirect bandgap, which results in low quantum efficiency for light emission—typically in the infrared region. One promising strategy to address this limitation is the development of nanocrystalline silicon, which consists of low-dimensional nanostructures such as quantum wires (QWs) and quantum dots (QDs). These structures exhibit enhanced photoluminescence and electroluminescence, primarily due to quantum confinement of excitons within the conduction and valence bands, leading to significantly improved quantum efficiency. Although research into these processes has spanned several decades, a definitive consensus on the mechanisms underlying photoluminescent emission in nanocrystalline silicon remains elusive. This work reviews two leading theoretical models proposed to explain this phenomenon and introduces a new, more comprehensive model that may provide a deeper and more accurate understanding of photoluminescence in nanocrystalline silicon. The new theory is applied in a case study.
\end{abstract}

\section*{}
The photoluminescence (PL) emitted from nanocrystalline objects within a porous silicon matrix has been studied extensively after its discovery by Canham \cite{Canham1990}. Since then, the scientific literature has defined some important bands classified according to typical PL decay lifetimes at room temperature \cite{Gelloz2018}, the most important being the S-band (wavelength from 400 to 1300 nm, blue to infrared), and the F-band (420 to 470 nm, blue to green).  Although there are some claims that PL comes from different mechanisms, it has been widely accepted that quantum confinement within nanocrystals (and with a possible contribution from surface states \cite{Hernández2015} ) is the principal way to explain the S-band, whereas surface chemistry plays a crucial role in explaining the F-band \cite{Saar2009}.

A few years after Canham´s seminal paper, John and Singh proposed a theory to explain the S-band in terms of distributions of spherical nanocrystals (named Quantum Dots, QD) and of nanocrystals in the form of columns (named Quantum Wires, QW) \cite{John1994}. In this paper, photoinduced exciton recombination (or band-to-band energy transition) in a nanostructure (QD or QW) of diameter D has the form:

\begin{equation} 
  \Delta E = c / D^2
\end{equation}

\noindent where c is a constant. By proposing a Gaussian distribution of diameters in the form:

\begin{equation} 
  P(D) = \frac {1}{{\sqrt {2\pi}}\sigma }
     \exp \left[ -\frac {1}{2} \left( \frac {D-D_0}{\sigma} \right)^2 \right]
\end{equation}

\noindent where $D_0$ is a mean diameter, and $\sigma$ is the standard deviation of diameters. For QW, the number of electrons participating in the exciton recombinations must be proportional to the wire surface (that is, equal to $a_W D^2$, with $a_W$ a constant), whereas for QD, it must be proportional to the sphere volume ($a_D D^3$). This gives rise to two components of the PL, whose line shapes must be calculated in the following forms:

\begin{equation} 
  \begin{split}
    \mathcal{P}_{QW} & (\Delta E) = \\
       & \frac {a_W}{{\sqrt {2\pi}}\sigma_W } \int_{0}^{\infty}
       (D_W)^2 \exp \left[ -\frac {1}{2} \left( \frac {D_W-D_{0,QW}}{\sigma_{QW}} \right)^2 \right]
       \delta \left( \Delta E - \frac {c}{(D_W)^2} \right)
       \,dD_W
  \end{split}
\end{equation}

\begin{equation} 
  \begin{split}
    \mathcal{P}_{QD} & (\Delta E) = \\
       & \frac {a_D}{{\sqrt {2\pi}}\sigma_D } \int_{0}^{\infty}
       (D_D)^3 \exp \left[ -\frac {1}{2} \left( \frac {D_D-D_{0,QD}}{\sigma_{QD}} \right)^2 \right]
       \delta \left( \Delta E - \frac {c}{(D_D)^2} \right)
       \,dD_D
  \end{split}
\end{equation}

\noindent The Delta functions assure that the PL energy (of the emitted photon $\hbar \omega$) coincides with the recombination energy given by:

\begin{equation} 
  \Delta E = \hbar \omega - \left( E_g - E_b \right)
\end{equation}

\noindent where $E_g \approx 1.17 \, eV$ is the silicon bulk gap energy and $E_b \approx 0.15 \, eV$ is the exciton binding energy. We see that: $\Delta E \approx \hbar \omega - 1 \, (eV)$. By changing variables in Eqs. (3) and (4) to energy, it is straightforward to find the PL components for QW and QD in the following forms:

\begin{equation} 
  \begin{split}
    \mathcal{P}_{QW} & (\Delta E) = \\
        & K_{QW} ( \Delta E )^{-5 / 2} 
         \exp \left[
                 -\frac {1}{2} \left( \frac {D_{0,QW}}{\sigma_{QW}} \right)^2
                   \left[ \frac {1}{D_{0,QW}} \left( \frac {c}{\Delta E} \right)^{1/2} - 1 \right]^2
              \right]
  \end{split}
\end{equation}

\begin{equation} 
  \begin{split}
    \mathcal{P}_{QD} & (\Delta E) = \\
        & K_{QD} ( \Delta E )^{-3} 
         \exp \left[
                 -\frac {1}{2} \left( \frac {D_{0,QD}}{\sigma_{QD}} \right)^2
                   \left[ \frac {1}{D_{0,QD}} \left( \frac {c}{\Delta E} \right)^{1/2} - 1 \right]^2
              \right]
  \end{split}
\end{equation}

\noindent where $K_{QW}$ and $K_{QD}$ are constants. Notice that for low values of the QW and QD standard deviations, the line profiles of Eqs. (6) and (7) are essentially Gaussian, with a possible skew at lower energies \cite{John1994}. The utility of these expressions is that one can extract nanostructure information from the PL spectrum of the S-band.

The cornerstone of the John-Singh theory lies in the validity of Eq. (2). Unfortunately, for silicon nanostructures, this is not the case. This problem was addressed by Elhouichet et al, a few years after John-Singh's theory proposal \cite{Elhouichet1997}. In their model, the Elhouichet group uses the following:

\begin{equation} 
  \Delta E = c / D^n
\end{equation}

\noindent with $n$ a value that usually has a value between 1 and 2. With this form for the exciton recombination energy, and reproducing the computations of John and Singh, one can obtain the following PL line forms:

\begin{equation} 
  \begin{split}
    \mathcal{P}_{QW} & (\Delta E) = \\
        & K_{QW} ( \Delta E )^{-(1 + 3/n)} 
         \exp \left[
                 -\frac {1}{2} \left( \frac {D_{0,QW}}{\sigma_{QW}} \right)^2
                   \left[ \frac {1}{D_{0,QW}} \left( \frac {c}{\Delta E} \right)^{1/n} - 1 \right]^2
              \right]
  \end{split}
\end{equation}

\begin{equation} 
  \begin{split}
    \mathcal{P}_{QD} & (\Delta E) = \\
        & K_{QD} ( \Delta E )^{-(1 + 4/n)} 
         \exp \left[
                 -\frac {1}{2} \left( \frac {D_{0,QD}}{\sigma_{QD}} \right)^2
                   \left[ \frac {1}{D_{0,QD}} \left( \frac {c}{\Delta E} \right)^{1/n} - 1 \right]^2
              \right]
  \end{split}
\end{equation}

\noindent Using the value given by Ref. \cite{Delerue1993}, where $n=1.39$, Eqs. (9) and (10) take the forms:

\begin{equation} 
  \begin{split}
    \mathcal{P}_{QW} & (\Delta E) = \\
        & K_{QW} ( \Delta E )^{-3.16} 
         \exp \left[
                 -\frac {1}{2} \left( \frac {D_{0,QW}}{\sigma_{QW}} \right)^2
                   \left[ \frac {1}{D_{0,QW}} \left( \frac {c}{\Delta E} \right)^{0.72} - 1 \right]^2
              \right]
  \end{split}
\end{equation}

\begin{equation} 
  \begin{split}
    \mathcal{P}_{QD} & (\Delta E) = \\
        & K_{QD} ( \Delta E )^{-3.88} 
         \exp \left[
                 -\frac {1}{2} \left( \frac {D_{0,QD}}{\sigma_{QD}} \right)^2
                   \left[ \frac {1}{D_{0,QD}} \left( \frac {c}{\Delta E} \right)^{0.72} - 1 \right]^2
              \right]
  \end{split}
\end{equation}

\noindent This model does not account for the participation of surface states on the nanocrystals. A good model that includes such states was proposed by Wolkin and coworkers \cite{Wolkin1999}, where a further component in the PL spectrum comes from exciton recombination in localized-to-band states. The origin of those states comes from the silicon-oxygen bonds present in nanocrystals smaller than 3 nm in diameter. A model containing the band-to-band recombinations (Elhouichet model) and the localized-to-band transitions (Wolkin model) was proposed by our group some time after the latter model was published \cite{RamirezPorras2002}. In this paper, we proposed a Gaussian distribution in localized exciton recombinations as follows:

\begin{equation} 
  \mathcal{P}_{loc}(\Delta E) = K_{loc}
     \exp \left[ -\frac {1}{2} \left( \frac {(\Delta E) - (\Delta E)_0}{\sigma_{loc}} \right)^2 \right]
\end{equation}

\noindent where $K_{loc}$ is a normalization constant, $(\Delta E)_0$ the distribution mean, and $\sigma_{loc}$ the standard deviation.

Eqs. (11), (12), and (13) are useful to extract information on nanostructures and surface status from PL measurements on porous silicon. But there is still a drawback. For conditions where the QW and QD standard deviations are high at the same time that their mean diameters are small, the Gaussian distribution of diameters would yield negative diameters at some of the values and, therefore, Eqs. (11) and (12) no longer hold. In such a case, it seems more plausible to propose a Logarithmic Normal distribution instead of a Gaussian distribution of nanostructure sizes. This kind of distribution has been previously proposed \cite{Yorikawa1997}, although it has not been applied in this theory.

To fix the computations, we propose the following logarithmic normal (LogNormal) distribution of diameters:

\begin{equation} 
  \mathcal{P}(D) = \frac {1}{{\sqrt {2\pi}}\sigma D}
     \exp \left[ -\frac {1}{2} \left( \frac {\ln(D)-\alpha}{\beta} \right)^2 \right]
\end{equation}

\noindent where $\alpha$ and $\beta$ are parameters related to the mean diameters and standard deviations in the following forms \cite{Dunn2023}:

\begin{equation} 
  D_0 =
    \exp \left( \alpha + \beta^2/2 \right) \, ,
    \, \sigma^2 = \left[ \exp \left( \beta^2 \right) -1 \right] \exp \left[ 2\alpha + \beta^2 \right]
\end{equation}

\noindent Figure 1 compares the line shapes of the LogNormal distribution function (Eq. (14)) and of the Gaussian distribution function (Eq. (2)) when $D_0=1.5 \, nm$ and $\sigma=0.8 \, nm$. Now it is clear that the LogNormal distribution is more convenient than the Gaussian one since the former has a zero value at zero diameter, whereas the latter does not vanish, which does not seem to have any physical meaning.

\begin{figure} 
    \centering
    \includegraphics[width=0.8\linewidth]{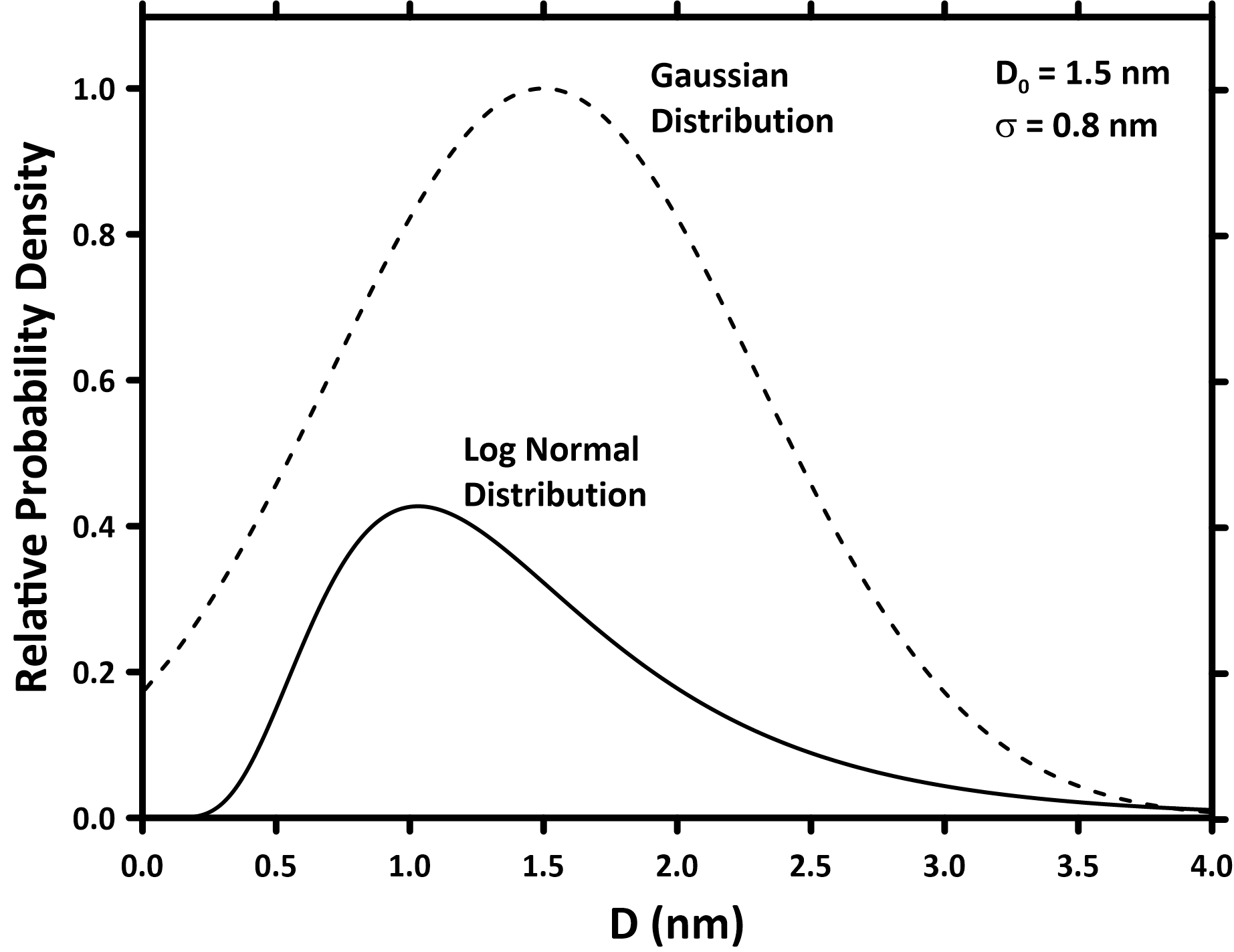}
    \caption{LogNormal (full line) and Gaussian (dotted line) distributions of nanocrystal sizes. Both curves were obtained using the same parameters $D_0$ and $\sigma$, as marked in the figure.}
    \label{fig:Fig1}
\end{figure}

With the LogNormal distributions of sizes given by Eq. (14) for both QW and QD, their line shapes must be calculated from the following:

\begin{equation} 
  \begin{split}
    \mathcal{P}_{QW} & (\Delta E) = \\
       & \frac {a_W}{{\sqrt {2\pi}}\sigma_W } \int_{0}^{\infty}
       D_W \exp \left[ -\frac {1}{2} \left( \frac {\ln (D_W)-\alpha_W}{\beta_W} \right)^2 \right]
       \delta \left( \Delta E - \frac {c}{(D_W)^n} \right)
       \,dD_W
  \end{split}
\end{equation}

\begin{equation} 
  \begin{split}
    \mathcal{P}_{QD} & (\Delta E) = \\
       & \frac {a_D}{{\sqrt {2\pi}}\sigma_D } \int_{0}^{\infty}
       (D_D)^2 \exp \left[ -\frac {1}{2} \left( \frac {\ln (D_D)-\alpha_D}{\beta_D} \right)^2 \right]
       \delta \left( \Delta E - \frac {c}{(D_D)^n} \right)
       \,dD_D
  \end{split}
\end{equation}

Once again, by changing variables and after performing the integrations, one obtains the following forms:

\begin{equation} 
  \begin{split}
    \mathcal{P}_{QW} & (\Delta E) = \\
        & K_W ( \Delta E )^{-(1 + 2/n)} 
         \exp \left[
              -\frac {1}{2\beta_W^2} \left( \frac {1}{n} \ln \left( \frac {c}{\Delta E} \right) - \alpha_W \right)^2
              \right]
  \end{split}
\end{equation}

\begin{equation} 
  \begin{split}
    \mathcal{P}_{QD} & (\Delta E) = \\
        & K_D ( \Delta E )^{-(1 + 3/n)} 
         \exp \left[
                 \frac {1}{2\beta_D^2} \left( \frac {1}{n} \ln \left( \frac {c}{\Delta E} \right) - \alpha_D \right)^2
              \right]
  \end{split}
\end{equation}

\noindent For $n=1.39$, we finally find:

\begin{equation} 
  \begin{split}
    \mathcal{P}_{QW} & (\Delta E) = \\
        & K_W ( \Delta E )^{-2.44} 
         \exp \left[
              -\frac {1}{2\beta_W^2} \left( 0.72 \ln \left( \frac {c}{\Delta E} \right) - \alpha_W \right)^2
              \right]
  \end{split}
\end{equation}

\begin{equation} 
  \begin{split}
    \mathcal{P}_{QD} & (\Delta E) = \\
        & K_D ( \Delta E )^{-3.16} 
         \exp \left[
                 -\frac {1}{2\beta_D^2} \left( 0.72 \ln \left( \frac {c}{\Delta E} \right) - \alpha_D \right)^2
              \right]
  \end{split}
\end{equation}

\noindent The incorporation of the Wolkin model is performed in the following way. Each of the nanostructures considered here must be passivated with a layer of oxide as proposed in Ref. \cite{Wolkin1999}. The localized-to-band recombinations must therefore originate from both the QW and QD (see Fig. 2 for an illustration of the case of a QD).

\begin{figure} 
    \centering
    \includegraphics[width=0.8\linewidth]{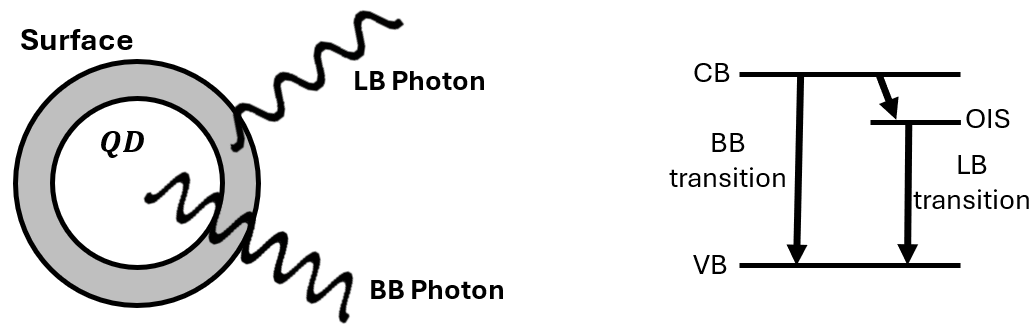}
    \caption{(Left) Interpretation of the photon emission of a QD. Two photons come from the inner nanocrystalline core due to a Band-to-Band (BB) transition, while a Localized-to-Band (LB) transition photon comes from the oxidized surface of the QD. (Right) The BB transitions are due to exciton recombinations from the conduction band (CB) to the valence band (VB). The LB transitions come from the oxide interface states (OIS) at the surface to the VB.}
    \label{fig:Fig2}
\end{figure}

\noindent We therefore propose, instead of just one localized exciton recombination (as shown in Eq. (13)), two distributions of exciton recombinations, one coming from the QW, and the other from the QD distributions, having each one a LogNormal form as follows:

\begin{equation} 
  \mathcal{P}_{loc}(\Delta E_{QW}) = K_{loc,QW}
     \exp \left[ -\frac {1}{2} \left( \frac {\ln \left[ (\Delta E_{QW}) / (\Delta E_{QW})_0 \right]}{\sigma_{loc,QW}} \right)^2 \right]
\end{equation}

\begin{equation} 
  \mathcal{P}_{loc}(\Delta E_{QD}) = K_{loc,QD}
     \exp \left[ -\frac {1}{2} \left( \frac {\ln \left[ (\Delta E_{QD}) / (\Delta E_{QD})_0 \right]}{\sigma_{loc,QD}} \right)^2 \right]
\end{equation}

\noindent where $K_{loc,QW}$ and $K_{loc,QD}$ are normalization constants, $(\Delta E_{QW})_0$ and $(\Delta E_{QD})_0$ the mean energy of the emitted photons and $\sigma_{loc,QW}$ and $\sigma_{loc,QD}$ the standard deviations of the distributions. Note that the emitted photons that come from the localized states in the oxide layers belong to their respective nanocrystals (QW and QD), and their mean energies $(\Delta E_{QW})_0$ and $(\Delta E_{QD})_0$ must be related to their respective nanocrystal mean diameter, namely, $D_{0,QW}$ and $D_{0,QD}$, as can be calculated from Eq. (15).

In summary, we visualize the PL spectrum as constructed from 4 components: band-to-band exciton recombination (Eqs. (20) and (21)), and localized-to-band recombinations (Eqs. (22) and (23)).

\begin{figure} 
    \centering
    \includegraphics[width=0.8\linewidth]{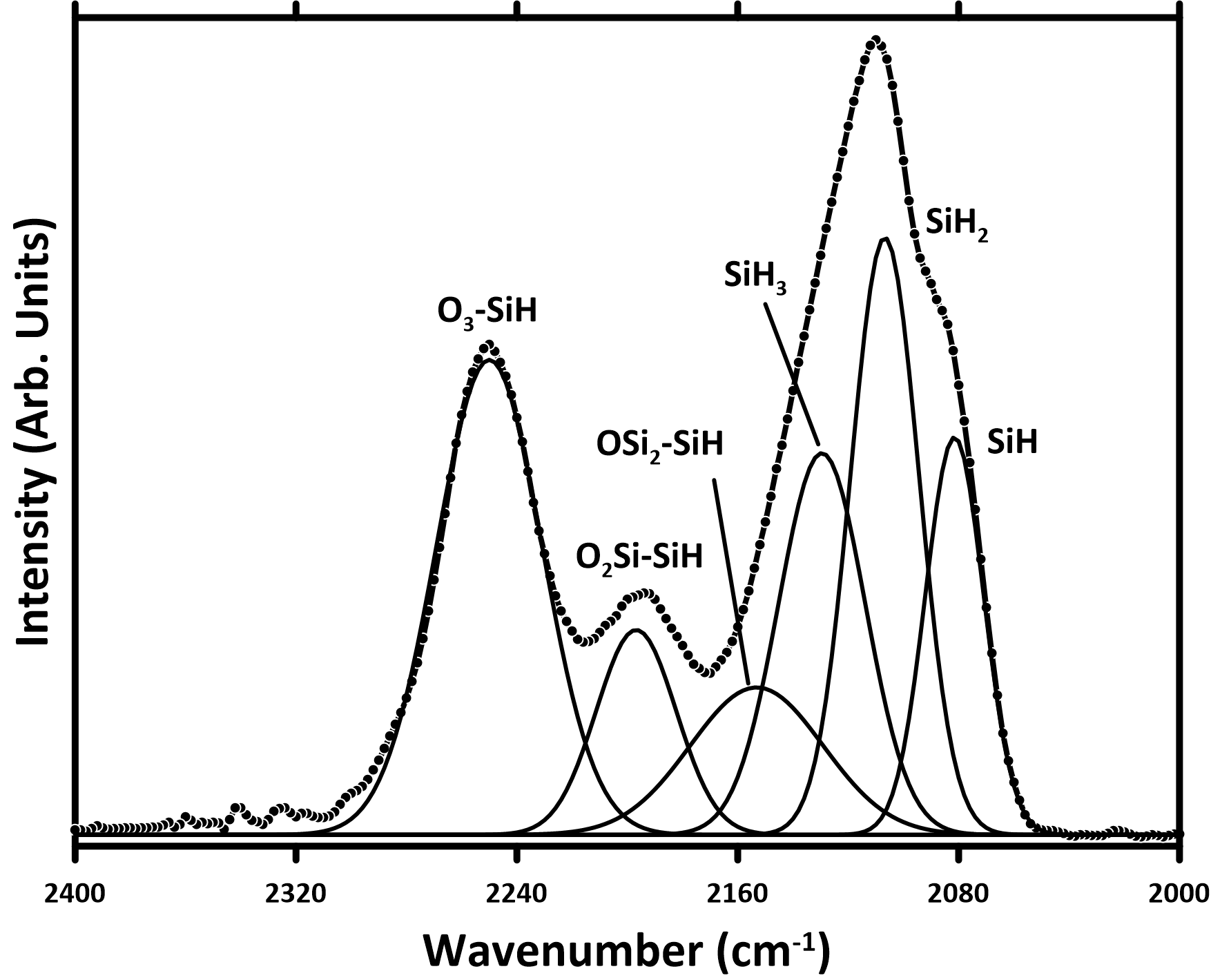}
    \caption{FTIR spectrum of a freshly synthesized porous silicon sample near the region of 2100 $cm^{-1}$. The dotted line represents the experimental data, and the continuous curves are the different silicon hydride bands (below 2150 $cm^{-1}$) and the silicon oxide bands (above 2150 $cm^{-1}$).}
    \label{fig:Fig3}
\end{figure}

Our new model was experimentally tested with a porous silicon film. We used a Teflon cell containing an ethanoic hydrofluoric acid solution with an acid concentration of 12.5\%. At the bottom of the cell, and in contact with the solution, a piece of boron-doped crystalline silicon of 1 cm x 1 cm size, resistivity of 10 –30 $\Omega \cdot$ cm, (100) crystal surface orientation, and 300 $\mu$ m thick. A Platinum electrode was immersed in the solution, and an Aluminum contact was placed in tight contact with the opposite side of the silicon crystal. A power source was used under galvanostatic conditions, providing a current density of 54 $mA/cm^2$ for 10 min, using the Al electrode as anode. The obtained porous layer had a thickness of 10 $\mu$m, 50\% porosity, and a total diameter of nearly 4 mm.

To check for the presence of Silicon oxide bands, a PerkinElmer Spotlight 400 Image Fourier Transform Infrared Spectrometer (FT-IR) was employed. The scanned sample area was 100 $\mu$m x 100 $\mu$m, with 32 scans per measurement. Fig. 3 shows the obtained spectrum in the region going from 2000 to 2400 $cm^{-1}$. A set of 6 peaks can be detected: three belonging to Silicon hydride species (SiHx, x=1, 2, 3) at 2081.5, 2106.8, and 2129.5 $cm^{-1}$, respectively, and the rest belonging to Silicon oxide species (OxSi3-x-SiH, x=1, 2, 3) at 2153.2, 2196.8, and 2250.0 $cm^{-1}$, respectively, corresponding closely with data provided by Ref. \cite{Ogata2018}. The silicon hydride bands passivate the silicon surface, whereas the silicon oxide bonds form when the semiconductor is placed in contact with the ambient air [8].

The PL spectrum was recorded using a PerkinElmer FL-8500 Fluorescence Spectrometer in emission mode (employing an excitation light source of 350 nm). A previously nontreated silicon surface spectrum was recorded and used as a reference to extract the actual PL spectrum. Fig. 4(a) shows the resulting PL spectrum, along with the fitted curves using Eqs. (20) to (23): band-to-band transitions within QW and QD (labeled as such in the graph), and localized-to-band transitions on the nanocrystals (dotted lines). The fittings were carried out using the Fityk program. The fitting parameters obtained from the models are presented in Table I. The estimated diameters of QW and QD are approximately 2 nm, with QW exhibiting slightly larger values. The energy transitions associated with localized states in both QW and QD are observed near 2 eV. These values are plotted in Fig. 4(b), where the plots from the Wolkin s et al. model \cite{Wolkin1999} are shown as full lines. As introduced in Ref. \cite{Wolkin1999}, three zones can be identified: Zone I (for nanostructures with diameters higher than 3 nm) does not exhibit localized-to-band transitions, Zone II (for diameters between 1.5 and 3 nm) where localized-to-band transition energies split out from the band-to-band transitions and tend to increase with decreasing nanostructure diameters (but with a slower rate than the band-to-band transitions), and Zone III (for diameters lower than 1.5 nm) where localized-to-band transition energies are almost constant for decreased diameters.

\begin{figure} 
    \centering
    \includegraphics[width=0.7\linewidth]{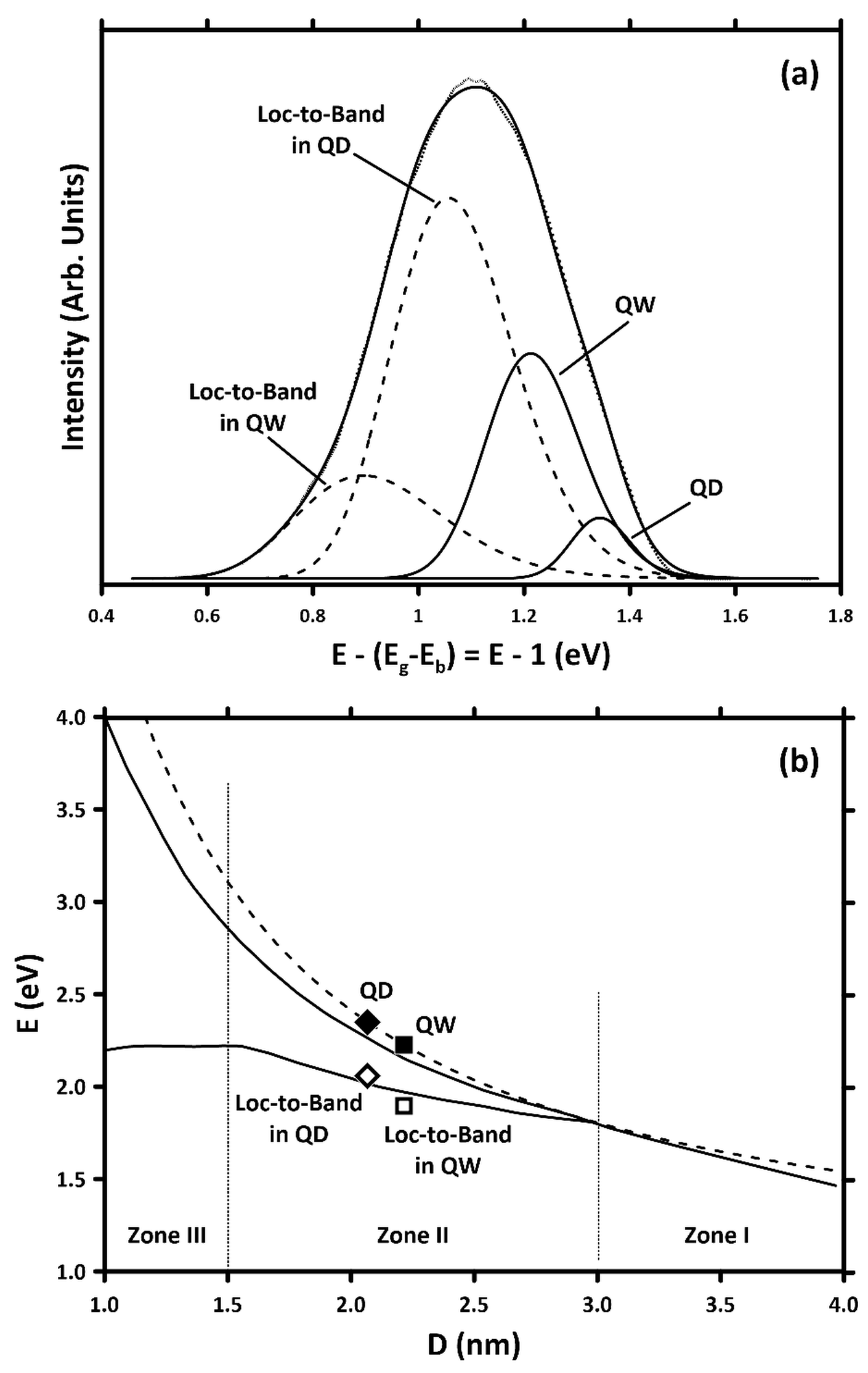}
    \caption{(a) PL obtained for the same sample whose IR spectrum was recorded and shown in Fig. 3. The experimental PL intensity (shown as small dots) is plotted against the recombination energy given in Eq. (5). Band-to-band transitions for QW and QD are shown as full lines, whereas Localized-to-band transitions in QW and QD are shown as dotted lines. (b) Plot of energy transitions as a function of nanostructure diameters using the model of Ref. [8] (full lines) and the model of Ref. \cite{Elhouichet1997} (dotted line). Values extracted from fittings and reported in Table I are plotted as points in the graph.}
    \label{fig:Fig4}
\end{figure}

The Elhouichet et al. model has been inserted into Fig. 4(b) as a dotted line, accounting for a different value of the exponent in the diameter as compared with the Wolkin et al. model, as mentioned previously. The data seem to fit better in the former model for band-to-band transitions within the nanostructures and fit appropriately in the localized-to-band transitions curve for the latter model.

\begin{table} 
 \caption{Fitting parameters obtained from plots of Fig. 3(a). Numbers in parentheses represent the respective standard deviations}
  \centering
  \begin{tabular}{ | m{2.5cm} | m{2.5cm}| m{3cm} | m{3cm} | } 
    \toprule
    \cmidrule(r){1-2}
    QW mean diameter (nm) & QD mean diameter (nm)  & Loc-to-Band mean energy in QW (eV) & Loc-to-Band mean energy in QD (eV) \\
    \midrule
    2.21 (0.12) & 2.07 (0.06) & 1.90 (0.16) & 2.06 (0.11)\\
    \bottomrule
  \end{tabular}
  \label{tab:table}
\end{table}

In summary, an improved model for explaining PL within nanostructured silicon was presented. This model gathers two previous models, one for band-to-band transitions on one side and another for localized-to-band transitions on the other side. The model predicts the existence of four contributions, which come in pairs from QW and QD.

\section*{Acknowledgments}
This work was supported by the Vicerrectoría de Investigación of the Universidad de Costa Rica.

\bibliographystyle{unsrt}  
\bibliography{Paper2025}

\end{document}